\documentclass[twocolumn,showpacs,amsmath,nofootinbib,amssymb]{revtex4}

\usepackage{epsfig}
\usepackage{bm}

\newcommand{\p}{\ensuremath{{\rm p}}}
\newcommand{\n}[1][]{\ensuremath{{^{#1}{\rm n}}}}
\newcommand{\Hy}[1][1]{\ensuremath{^{#1}{\rm H}}}
\newcommand{\D}{\ensuremath{{\rm D}}}
\newcommand{\T}{\ensuremath{{\rm T}}}
\newcommand{\He}[1][4]{\ensuremath{^{#1}{\rm He}}}
\newcommand{\Li}[1][7]{\ensuremath{^{#1}{\rm Li}}}
\newcommand{\Be}[1][7]{\ensuremath{^{#1}{\rm Be}}}

\begin{document}
\title{The Effect of Bound Dineutrons upon BBN}

\author{James P. Kneller}
\email{Jim_Kneller@ncsu.edu} \affiliation{Department of Physics,
North Carolina State University, Raleigh, North Carolina
27695-8202}

\author{Gail C. McLaughlin}
\email{Gail_McLaughlin@ncsu.edu} \affiliation{Department of
Physics, North Carolina State University, Raleigh, North Carolina
27695-8202}

\date{\today}

\begin{abstract}
We have examined the effects of a bound dineutron, \n[2], upon big
bang nucleosynthesis (BBN) as a function of its binding energy
$B_{\n[2]}$. We find a weakly bound dineutron has little impact
but as $B_{\n[2]}$ increases its presence begins to alter the flow
of free nucleons to helium-4. Due to this disruption, and in the
absence of changes to other binding energies or fundamental
constants, BBN sets a reliable upper limit of $B_{\n[2]} \lesssim
2.5\;{\rm MeV}$ in order to maintain the agreement with the
observations of the primordial helium-4 mass fraction and \D/\Hy[]
abundance.
\end{abstract}

\pacs{99.99}

\keywords{BBN,nuclear physics,dineutron,binding energy}

\maketitle


\newpage

\section{Introduction} \label{sec:intro}

The concordance of the predicted synthesis of the lightest nuclei
during the period immediately following the Big Bang with the
observed primordial abundances presents us with a powerful probe
of the state of the Universe during its earliest epochs. In
addition to constraining standard cosmological parameters such as
the density of baryons, the number of neutrino flavors and their
degeneracy
\cite{S1998,VFCC2003,SSG1977,YSSR1979,LSV1999,KSSW2001,BY1977,KS1992,KKS1997,OSTW1991,S1983,WFH1967,YB1976,Betal2003a,Betal2003b}
BBN is sufficiently complex that one can also learn about such
possibilities as Quintessence
\cite{SDT1993,SDTY1992,BHM2001,KS2003}, modifications of gravity
\cite{KS2003,CSS2001} or neutrino oscillations/mass/decay
\cite{F2000,LS2001,Detal2002,W2002,ABB2002,KSW2000,KS1982,SSF1993,SFA1999}.

Perhaps the most intriguing use of BBN is in constraining the
variation of the fundamental constants
\cite{Bergstrom:1999wm,Aetal01,Nollett:2002da,Yoo:2002vw,Flambaum:2002de,Flambaum:2002wq,KM2003}.
Support for this hypothesis has emerged from recent observations
of quasar absorption lines at redshift of z=1-2 by Webb et al.
\cite{Webb:1998cq,Webb:2000mn} that suggest the fine structure
constant, $\alpha$, may have been smaller in the past (though see
\cite{Bahcall:2003rh}). In some cases the variation of a
fundamental constant is easily implemented in BBN because the
nuclear physics aspects of the calculation are unaffected, but in
others the lack of an adequate theory to predict such parameters
as the nuclear binding energies and cross sections introduces a
degree of uncertainty. An example is the calculation of Kneller \&
McLaughlin \cite{KM2003} who looked at the variation of
$\Lambda_{QCD}$, its effects upon the binding energy of deuterons
and the neutron-proton mass difference and, in turn, the impact
upon BBN. In addition to variation of these nuclear parameters,
variation of the constants relevant to nuclear structure might
also partially stabilize nuclei that are presently particle
unstable. For example, the lack of stable $A=5$ and $A=8$ nuclei
is often cited as the explanation for the dearth of nuclei formed
with masses above helium-4 though the endothermicity of pure
strong reactions such as $\He(\T,\n)\Li[6]$, $\He(\He,\p)\Li$ and
$\He(\He,\n)\Be$ plays a role.

The focus in this paper is upon another nucleus that could also
become stabilized - the dineutron \n[2]. The dineutron, a member
of the nucleon-nucleon isospin triplet, is a spin singlet and, by
itself, the dineutron is weakly unbound\footnote{Though it may
become stable on the surface of neutron-rich nuclei.} by $\sim
70\;{\rm keV}$, the \n-\n~ scattering length being negative
\cite{PC54,IKPS61,GGS1987,Setal87}. Early direct searches
\cite{CH53} did not see evidence for a stable dineutron but
recently Bochkarev {\it et al.} \cite{Betal1985a} claim that
$45\pm 10$\% of the decay of an excited state of \He[6] is through
the dineutron state \footnote{Bochkarev {\it et al.}
\cite{Betal1985b} also have evidence of diproton emission from an
excited state of \Be[6].} and Seth \& Parker \cite{SP1991},
amongst others, find evidence for dineutrons in \Hy[5], \Hy[6] and
\He[8] decay. There is also a claim for tetraneutron, \n[4],
emission in the decay of \Be[14] \cite{Metal2003}.

Given that the dineutron is only weakly unbound, even small
changes in the pion mass could, perhaps, result in a bound
dineutron, although at present there is not enough experimental
information to show whether or not this would occur
\cite{Beane:2002vq,Beane:2002xf}. The scattering length is quite
sensitive to the pion mass and so it is small changes in
fundamental constants that change its mass, such as $\alpha$,
$\Lambda_{QCD}$ or the Higgs vev, that could cause the dineutron
to become bound. Since we are lacking an exact relationship
between these fundamental constants and the binding energy of the
dineutron we do not adopt a particular model for the time
variation of fundamental constants, but instead explore the effect
upon BBN of a bound dineutron directly.

In this paper we shall make an effort to derive a constraint upon
the dineutron binding energy. We will consider the dineutron in
isolation i.e. whatever the source of the new stability of the
dineutron we shall limit the effect to just this nucleus. We begin
with an overview of standard BBN in section \S\ref{sec:SBBN} with
an emphasis on the details of the flow from free nucleons to
helium-4 before proceeding to insert dineutrons in section
\S{\ref{sec:BBN+n2}. In section \S{\ref{sec:BBN+n2 results} we
present our results for a baryon-to-photon ratio of $\eta = 6.14
\times 10^{-10}$ and follow it up in \S\ref{sec:MC} with a
discussion of the errors in the calculation and any degeneracy
with $\eta$ in \S\ref{sec:eta}. Finally, in \S\ref{sec:limits}, we
show how BBN can limit the dineutron binding energy before
presenting our conclusions.


\section{BBN without dineutrons.} \label{sec:SBBN}

BBN can be simplistically broken into three phases characterized
by the degree of equilibrium within the nucleons/nuceli.

During the first phase, at temperatures above $T \gtrsim 1\;{\rm
MeV}$, there are virtually no complex nuclei so that all the
nucleons exist in a free state. The rapidity of the weak
interactions in converting neutrons and protons,
\begin{subequations}
\begin{eqnarray}
\n & \leftrightarrow & \p + e + \bar{\nu}_{e} \label{reac:ndecay}\\
\n+\bar{e} & \leftrightarrow & \p + \bar{\nu}_{e}, \label{reac:nebar}\\
\n+\nu_{e} & \leftrightarrow & \p + e, \label{reac:nnue}
\end{eqnarray}
\end{subequations}
establish a weak equilibrium so that the neutron/proton ratio,
$F$, is simply $F \approx \exp( -\Delta_{\n\p}/T)$. As the
Universe cools the rate at which neutrons must be converted to
protons in order to maintain the equilibrium cannot be
accommodated. As a consequence, the neutron-to-proton ratio
departs from its equilibrium value and is said to `freeze-out'
even though conversion continues to occur. In the absence of
neutron decay and the formation of complex nuclei, the ratio would
attain an asymptotic value of $F\sim 1/6$ \cite{BBF1989,S1996}.
The departure of $F$ from its equilibrium value denotes the
boundary between the first two phases of BBN.

During the second phase of BBN, from a temperature of $\sim
1\;{\rm MeV}$ to $\sim 100\;{\rm keV}$, the abundances of the
various nuclei also begin to depart from equilibrium. At $\sim
1\;{\rm MeV}$ their abundances are suppressed relative to the free
nucleons but the nuclear reactions that form them establish, and
maintain, chemical/nuclear statistical equilibrium (NSE). In
equilibrium the abundance\footnote{The term `abundance' is also
used for the ratio $Y_{A}/Y_{H}$}, $Y_{A}=n_{A}/n_{B}$, of a
complex nuclei $A$ is derived from $\mu_{A} = Z\,\mu_{\p} +
(A-Z)\,\mu_{\n}$ so after inserting the expressions for the
Boltmann number density we find
\begin{equation}
Y_{A} = \frac{g_{A}\,A^{3/2}}{2^{A}}\, \left[ n_{B}\,
\left(\frac{2\,\pi}{m_{N}\,T}\right)^{3/2} \right]^{A-1}
\,Y_{\p}^{Z}\,Y_{\n}^{A-Z}\, e^{B_{A}/T}. \label{eq:NSE}
\end{equation}
Using $F\sim 1/6$ and a baryon-photon ratio of $\eta \sim
10^{-10}$ we see that for a temperature of $T \sim 1\;{\rm MeV}$
the abundance of deuterons is $Y_{\D} \sim 10^{-12}$. After
substituting $Y_{\D}$ for the thermal factors in equation
(\ref{eq:NSE}) we obtain
\begin{eqnarray}
Y_{A} & = & \frac{g_{A}\,A^{3/2}}{2\,[3\,\sqrt{2}]^{A-1}}\,Y_{\p}^{1+Z-A}\,Y_{\n}^{1-Z}\,Y_{\D}^{A-1} \nonumber \\
 & & \times \exp \left( \frac{B_{A}-(A-1)B_{\D}}{T} \right).
\end{eqnarray}
This equation now makes it much clearer that the abundance of a
nucleus with mass $A+1$ is suppressed by approximately $Y_{\D}$
relative to the abundance of a nucleus with mass $A$.

If the neutron/proton abundances are held fixed then the NSE
abundance has a minimum at $T_{A} = 2\,B_{A}/( 3A-3)$. Below
$T_{A}$ the various nuclear reactions provide sufficient nuclei to
keep $Y_{A}$ in equilibrium as the abundance climbs from the
minimum but eventually a point is reached where this required
production rate cannot be met and the abundance falls beneath the
equilibrium value. This departure from equilibrium for the complex
nuclei is in contrast with that of the neutrons where it was an
insufficient rate that led to the departure, here it is a lack of
reactants that is the cause. Heavier nuclei are the first to
depart from equilibrium: helium-4 departs at $T \sim 600\;{\rm
keV}$ while helium-3 and tritium drop out at $T\sim 200\;{\rm
keV}$. Below $T \sim 200\;{\rm keV}$ the only compound nucleus in
NSE is the deuteron and its abundance controls the rate at which
all the heavier nuclei can be produced. By $T \sim 100\;{\rm keV}$
the \D~ abundance is approaching that of the free nucleons and the
amount of \D~ destruction, via such reactions as $\D(\D,\p)\T$ and
$\D(\D,\n)\He[3]$, has becomes significant. When this occurs the
deuteron abundance cannot be replenished sufficiently quickly to
maintain its equilibrium and, consequently, it too departs from
NSE. This final NSE departure forms the entrance to the third
stage of BBN proper.

Below the deuteron NSE departure temperature the \D~ abundance
continues to grow for a short period but eventually the $\D+\D$
drain tips the balance in favor of destruction and the deuteron
abundance reaches a peak amplitude. The tritons and helions formed
via the reactions $\D(\D,\p)\T$ and $\D(\D,\n)\He[3]$ are produced
in roughly equal amounts but the helions rapidly transform to
tritium via $\He[3](\n,\p)\T$. The last step in the formation of
the alpha particle is almost exclusively $\T(\D,\n)\He$ which
destroys $\sim 1/3$ of all the deuterons formed. The essential
steps in the scheme are illustrated in figure (\ref{flow:SBBN}).
\begin{center}
\setlength{\unitlength}{1.5cm}
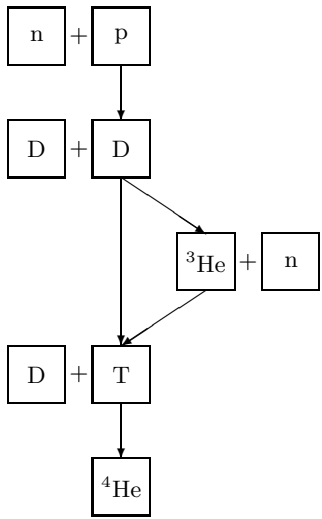
\begin{figure}[htbp]
\begin{picture}(2.75,4.5)(0,0)
\put(0,4){\framebox(0.5,0.5){\n}}
\put(0.5,4){\makebox(0.25,0.5){+}}
\put(0.75,4){\framebox(0.5,0.5){\p}}
\put(0,3){\framebox(0.5,0.5){\D}}
\put(0.5,3){\makebox(0.25,0.5){+}}
\put(0.75,3){\framebox(0.5,0.5){\D}}
\put(1.5,2){\framebox(0.5,0.5){\He[3]}}
\put(2,2){\makebox(0.25,0.5){+}}
\put(2.25,2){\framebox(0.5,0.5){\n}}
\put(0,1){\framebox(0.5,0.5){\D}}
\put(0.5,1){\makebox(0.25,0.5){+}}
\put(0.75,1){\framebox(0.5,0.5){\T}}
\put(0.75,0){\framebox(0.5,0.5){\He}}
\put(1,4){\vector(0,-1){0.5}}
\put(1,3){\vector(0,-1){1.5}}
\put(1,3){\vector(3,-2){0.75}}
\put(1.75,2){\vector(-3,-2){0.75}}
\put(1,1){\vector(0,-1){0.5}}
\end{picture}
\caption{A diagrammatic flow of nuclei in standard BBN. The
complex nuclei are immersed in a (rapidly drained) bath of free
nucleons and so we outline only where they form part of the set of
reactants. \label{flow:SBBN}}
\end{figure}
\end{center}
Smith, Kawano and Malaney \cite{Smith:1992yy} identified 8
reactions among the $A \leq 4$ nuclei as being important for BBN.
Five are identical to those in figure (\ref{flow:SBBN}), the sixth
is neutron-proton interconversion, and the two reactions they
included, and which we have omitted from the figure, are
$\He[3](\D,\p)\He$ and $\T(\p,\gamma)\He$ both of which are $2-4$
orders of magnitude smaller than $\T(\D,\n)\He$, so there
importance is marginal. Whatever the exact number, this handful
(or two) of important reactions is much smaller than the number of
reactions included in any BBN code.

As the Universe cools eventually the Coulomb barriers in the
various reactions become insurmountable leading to a cessation of
the nucleosynthesis. The abundances have plateaued to their
`primordial' values with virtually every neutron now incorporated
in helium-4 and only small residues in \D, \T~ and \He[3]. A small
(but detectable) abundance of lithium-7 and beryllium-7 have also
been formed but we will not discuss these two nuclei further.


\section{Inserting dineutrons into BBN} \label{sec:BBN+n2}

Inclusion of a new, light nucleus into BBN has the potential to
significantly influence BBN and alter the predicted primordial
abundances. These changes will occur because the dineutron will
disrupt the flow of nucleons through the reaction network by both
presenting new exit channels for reactions that already exist in
BBN and through new entrance channels in the formation of the
nuclei. One can construct a large number of plausible reactions in
which dineutrons participate but not all are expected to play a
prominent role for the same reason that standard BBN is dominated
by only a few reactions. We can use our understanding of the
important reactions in standard BBN to pick from the plethora of
possibilities for dineutron reactions those which we expect to be
important. The most important reactions involving dineutrons
should be
\begin{itemize}
  \item preferably exothermic,
  \item dominated by the strong interaction,
  \item and two-bodied in their entrance channel.
\end{itemize}
Exothermicity plays a pivotal role in the BBN because, during its
second two phases, the system does not attain an equilibrium and,
typically, the flow of nuclei in any given reaction is one
direction only. In a few cases where the Q-value for the reaction
is less than few times the temperature during BBN (i.e. $T \sim
100\;{\rm keV}, Q\sim 500\;{\rm keV}$) flow may be in both
directions because the `activation energy' for an endothermic
reaction is readily available. Examples of this behavior were seen
in Kneller and McLaughlin \cite{KM2003}. The reaction should also
be preferably strong in nature since this is the behavior seen in
standard BBN where reactions such as $\D(\D,\n)\He[3]$ dominate
over $\D(\p,\gamma)\He[3]$ though, in a few cases, such as
$\p(\n,\gamma)\D$, electromagnetic or weak interactions play an
important role. The last requirement will remove those cases that
pass the first two but whose entrance channels involve multiple
particles.

From these requirements we have selected four dineutron reactions
that we expect to be important:
\begin{eqnarray}
\p+\n[2] \leftrightarrow \D +\n , & & Q = 2.22\;{\rm MeV} - B_{\n[2]}, \label{reac:n2+p} \\
\D+\n[2] \leftrightarrow \T + \n , & & Q = 6.26\;{\rm MeV} -
B_{\n[2]}, \label{reac:d+n2}
\end{eqnarray}
\begin{subequations}
\begin{eqnarray}
\He[3]+\n[2] \leftrightarrow \He+\n ,  & & Q = 28.29\;{\rm MeV} - B_{\n[2]}, \label{reac:he3+n2->he4n} \\
\He[3]+\n[2] \leftrightarrow \T + \D , & & Q = 2.99\;{\rm MeV} -
B_{\n[2]}. \label{reac:he3+n2->td}
\end{eqnarray}
\end{subequations}
The dineutron binding energy will determine the Q-value in these
reactions and, if we permit values of $B_{\n[2]}$ of several MeV,
both (\ref{reac:n2+p}) and (\ref{reac:he3+n2->td}) may reverse
sign. We will not consider any changes to the other nuclear
binding energies. Though one may expect a large change in the
dineutron binding energy to be reflected in equally significant
changes to the structure of the deuteron, the effects upon three
nucleon nuclei may be considerably smaller \cite{braaten}. In our
study, we also do not consider reactions leading to the
formation of destruction of nuclei above mass 4.

To this list we add three additional reactions:
\begin{eqnarray}
\n+\n \leftrightarrow \n[2]+ \gamma , & & Q = B_{\n[2]}, \label{reac:2n->n2}\\
\n[2] \leftrightarrow \D, & & \label{reac:n2->d} \\
\n[2]+\p \leftrightarrow \T +\gamma , & & Q = 8.48\;{\rm MeV} -
B_{\n[2]}. \label{reac:n2+p->t}
\end{eqnarray}
These reactions could become important for producing and, more
importantly, removing dineutrons. In particular, the inclusion of
(\ref{reac:n2->d}) and (\ref{reac:n2+p->t}) is based on the
following reasoning. Once we have dineutron cross sections we must
integrate them over a Maxwell-Boltzmann spectrum to obtain a
thermally averaged rate \cite{FCZ1967}. In all the dineutron
reactions the lack of a Coulomb barrier means that the cross
section for an exothermic reaction varies as $1/\sqrt{E}$ at low
energy and so the rate becomes a constant. This can have important
consequences: written in terms of the temperature, a reaction such
as $A+B\rightarrow X$, with a rate per particle pair $\Gamma$,
destroys nucleus $A$ at a rate
\begin{equation}
\frac{dY_{A}}{dT} \propto \Gamma\,Y_{A}\,Y_{B}
\label{eq:neverending BBN}
\end{equation}
where we have used the relations $n_{B} \propto T^{3}$ and $T^{2}
\propto 1/t$. If we assume the abundance of nucleus $B$ is much
larger than $A$'s and does not change by any other process then
the solution to this equation is $Y_{A} \propto
\exp(\Gamma\,Y_{B0}\,T)$ where $Y_{B0}$ is the abundance of $B$ at
some fiducial temperature. In this scenario, the abundance of $A$
never becomes a constant and BBN would never end! In standard BBN
this situation never arises because the two
temperature-independent reactions, $\p(\n,\gamma)\D$ and
$\He[3](\n,\p)\T$, are killed by the decay of the neutron. But if
the dineutron becomes stable
then without a reaction such as $\n[2](\p,\gamma)\T$ the dineutron
abundance could plateau to a constant larger than, say, the
abundance of helium-3 and the circumstances of equation
(\ref{eq:neverending BBN}) may be realized in the reaction
$\He[3](\n[2],\n)\He$. In order to obtain a primordial abundance
of $^3{\rm He}$ it is imperative that this situation be prevented.
The decay of the dineutron and $\n[2](\p,\gamma)\T$ will ensure
this by depleting the final dineutron abundance for all values of
$B_{\n[2]}$ we shall explore.

Now that we have determined the most important reactions, we need
cross sections for them before we can proceed.


\subsection*{$\n[2](\p,\n)\D$, $\D(\n[2],\n)\T$, $\He[3](\n[2],\n)\He$ and $\He[3](\n[2],\D)\T$}

Since the dineutron is presently unstable we posit cross sections
based on `similar' strong reactions involving deuterium or other
light nuclei. If we consider an arbitrary two-body strong reaction
$i+j \leftrightarrow k + l$ then general considerations lead us to
expect a cross section per particle pair which is proportional to
a matrix element squared, the phase space with an energy
conserving delta function and inversely proportional to a flux. In
nuclear astrophysics, one typically writes the cross section as
\begin{equation}
\sigma(E) = \frac{S(E)}{E}\; \exp\left( -\pi\,\alpha\,Z_{i}\,Z_{j}
\sqrt{\frac{2\,\mu_{ij}}{E}} \right) \label{eq:astroS}
\end{equation}
where $\mu_{ij}$ is the reduced mass for incoming particles $i$
and $j$, and $S(E)$ is the astrophysical S-factor. For our
purposes, this parameterization is not sufficient since it does
not explicitly show the effects of a Q value upon the final
states. This is particularly crucial since in our study Q-values
will vary as the dineutron binding energy varies. With that in
mind, we write the non-resonant (S-wave) contributions to the
cross section, following \cite{Gamow, Bethe}, as proportional to
the product of both Coulomb penetrability factors $G_{ij}(E)$ and
$G_{kl}(E+Q)$, the available phase space in the exit channel
$\Phi_{kl}(E+Q)$ together with the statistical weight $g_{kl}$ and
the reciprocal of the entrance channel velocity. The penetrability
factor for charged particle interactions, $G_{ij}(E)$, is simply
\begin{equation}
G_{ij}(E) = \sqrt{\frac{E_{ij}^{C}}{E}}\,\exp\left(
-\pi\,\alpha\,Z_{i}\,Z_{j} \sqrt{\frac{2\,\mu_{ij}}{E}} \right)
\end{equation}
where $E_{ij}^{C}$ is the Coulomb barrier energy \cite{Clayton83}.
Although in the standard cross section parameterization, shown in
equation (\ref{eq:astroS}), the second Coulomb penetrability
factor can be absorbed into the astrophysical S factor because,
typically, the energy is much smaller than the Q-value, here we
retain it explicitly. The phase space factor, $\Phi_{kl}(E+Q)$, is
\begin{equation}
\Phi_{kl}(E+Q) \propto \sqrt{ (E+Q)\,\mu^{3}_{kl} }
\end{equation}
while the statistical weight factor $g_{kl}$ accounts for the
multiplicity of the final state
\begin{equation}
g_{kl} = (2\,J_{k} +1 )(2\,J_{l}+1)
\end{equation}
where $J_{k}$ and $J_{l}$ are the spins of the individual nuclei.
The reciprocal of the entrance channel velocity is proportional to
$\Phi_{ij}/(\mu_{ij}E)$. Putting all these together the cross
section for the reactions $i+j \leftrightarrow k + l$ is expected
to be of the form
\begin{eqnarray}
\sigma_{i+j \rightarrow k + l}(E) & = &
S_{ij,kl}(E)\,\frac{g_{kl}}{\mu_{ij}\,E}\,G_{ij}(E)\,G_{kl}(E+Q)
\nonumber \\
& & \times \Phi_{ij}(E)\,\Phi_{kl}(E+Q)\, \label{eq:csNR}.
\end{eqnarray}
where we have introduced $S_{ij,kl}(E)$ as an undetermined
function. This function becomes constant at low energy, and is
similar to, but not the same as, the astrophysical S-factor.

With the help of equation (\ref{eq:csNR}) we can extract the most
obvious behavior of any cross section with energy to derive
$S_{ij,kl}(E)$. Our expectation is that this quantity varies
slowly though of course the exact details of a reaction may lead
to significant departures. After extracting $S_{ij,kl}(E)$ from
some known reaction we can then insert it into the similar
dineutron reaction based on the assumption that $S_{ij,kl}$ does
not change considerably from one to the other. The major changes
in the cross sections will therefore be limited to the
considerable effects of the Coulomb barrier penetrability and
phase space. We have examined the validity of this approach by
using it to predict $\D(\D,\p)\T$ from $\D(\D,\n)\He[3]$, and
$\T(\D,\n)\He$ from $\He[3](\D,\p)\He$. The last two cross section
are dominated by large resonances corresponding to excited states
of \He[5] and \Li[5] \cite{TUNL} but the non-resonant pieces have
been extracted by Chulick \emph{et al.} \cite{Cetal93} allowing us
to compare the transformation. In both test cases we find the
transformation works reasonably well with an error
that is a factor of order a few.

For the $\n[2](\p,\n)\D$ reaction there is no similar deuteron
reaction with which to compare so instead we used the broadly
similar $\He[3](\n,\p)\T$. We have been unable to find an analytic
expression for the this cross section so we interpolated the
ENDF-IV evaluated cross section data available online \cite{endf}
and then factored out the expected behavior shown in equation
(\ref{eq:csNR}) before replacing it with the appropriate terms for
$\n[2]+\p \leftrightarrow \D +\n$.

We expect the $\D+\n[2] \leftrightarrow \T + \n$ cross section to
be similar to $\D(\D,\p)\T$ and $\D(\D,\n)\He[3]$ up to
corrections for the Coulomb barrier penetrability and phase space
factors. Here we have analytic expressions of the S-factor to use
from Chulick \emph{et al.} \cite{Cetal93}.

To estimate the last two reactions, $\He[3]+\n[2] \leftrightarrow
\n + \He$ and $\He[3]+\n[2] \leftrightarrow \D + \T$, one appeals
to their similarity with $\T(\D,\n)\He$ so that one may use the
Chulick \emph{et al.} \cite{Cetal93} expression and, once again,
correct for the change in the phase space, Coulomb barrier
penetrability etc. As mentioned earlier, the $\T(\D,\n)\He$ cross
section exhibits a resonance due to an excited state of the \He[5]
nucleus (see \cite{TUNL} for an energy level diagram). The
position of this same resonance, relative to the $\He[3]+\n[2]$
ground state, depends on the dineutron binding energy being
subthreshold for $B_{\n[2]} \lesssim 3\;{\rm MeV}$. In addition,
there are further excited states of \He[5] that become relevant
when $B_{\n[2]} \sim 0$ but, as we will show, the effects of the
dineutron become apparent only when $B_{\n[2]}$ approaches
$B_{\D}$ and we have not added them to our cross section.


\subsection*{$\n[2] \leftrightarrow \D$}

This weak reaction is actually the sum of \emph{four}
sub-processes:
\begin{subequations}
\begin{eqnarray}
\n[2] \leftrightarrow \D + e + \bar{\nu}_{e} & & 0 \leq E_{\nu} \leq \Delta_{\n[2]\D}-m_{e} \label{reac:n2decay} \\
\n[2] + \nu_{e} \leftrightarrow \D + e       & & - \Delta_{\n[2]\D} + m_{e} \leq E_{\nu} \label{reac:n2nue} \\
\n[2] + \bar{e} \leftrightarrow \D + \bar{\nu}_{e}, & & \Delta_{\n[2]\D} + m_{e} \leq E_{\nu} \label{reac:n2ebar} \\
\n[2] + \bar{e} + \nu_{e} \leftrightarrow \D & & 0 \leq E_{\nu}
\leq -\Delta_{\n[2]\D} - m_{e} \label{reac:n2ebarnue}
\end{eqnarray}
\end{subequations}
where we have denoted by $\Delta_{\n[2]\D}$ the dineutron-deuteron
mass difference i.e $\Delta_{\n[2]\D} = \Delta_{\n\p}+B_{\D} -
B_{\n[2]} = 3.52\;{\rm MeV} - B_{\n[2]}$. We do not consider those
cases where the final states are free nucleons. From the limits on
the neutrino energy, $E_{\nu}$, we see that, at most, only three
of these reactions can be operant at any given value of
$B_{\n[2]}$. For these rates we use expressions similar to the
neutron-proton interconversion rates but with different Q-values
and a matrix element that takes into account the presence of two
neutrons.  We also use a pure Gamow-Teller decay between the
$0^{+}$ ground state of the dineutron and the $1^{+}$ ground state
of the deuteron. While there remains some uncertainty in the rates
it has a much smaller impact on final abundance yields as compared
with the uncertainties in the strong interaction rates discussed
above.

There is one interesting quirk that appears when $\Delta_{\n[2]\D}
< m_{e}$ , which is that an atom of deuterium can capture an
electron to form a dineutron. Although interesting this process
has not been included in our calculations because the amount of
atomic deuterium is negligible at the temperatures relevant to
BBN. We also see that spontaneous deuteron decay can occur when
$\Delta_{\n[2]\D} < -m_{e}$ i.e. when $B_{\n[2]} \gtrsim 4\;{\rm
MeV}$. Though this possibility remains interesting we shall not
pursue dineutron binding energies that permit this reaction to
occur.


\subsection*{$\n(\n,\gamma)\n[2]$ and $\n[2](\p,\gamma)\T$}

Though $\n(\n,\gamma)\n[2]$ looks similar to deuteron formation
via $\n(\p,\gamma)\D$ this reaction is suppressed because there is
no charge.
Despite this smallness, it is the only exothermic reaction capable
of producing dineutrons when $B_{\n[2]}$ is small. Following Rupak
\cite{Rupak:1999rk}, the lowest order contribution to this cross
section should be something like $(N^T \sigma_2 \otimes  \tau_2
\tau_j N)^\dagger (N^T \sigma_2 \otimes \tau_2 \tau_j (\buildrel
\leftharpoonup \over D_k + \buildrel \rightharpoonup \over D_k) N)
E_k$ and $(N^T \sigma_2 \otimes \tau_2 \tau_j N)^\dagger (N^T
\sigma_2 \sigma_i \otimes \tau_2 \tau_j (\buildrel \leftharpoonup
\over D_k - \buildrel \rightharpoonup \over D_k) N) B_l
\epsilon^{ikl}$. From examining the $\n(\p,\gamma)\D$ operators in
the cross section from Rupak we estimated that this lowest order
contribution is N$^{4}$LO. We therefore make an order of magnitude
estimate for the dineutron cross section by starting with the
$\n(\p,\gamma)\D$ (which has a leading order contribution) and
suppressing it by the appropriate factor. Although there is
considerable error, just as with the two $\He[3]+\n[2]$ reactions,
the effects of the dineutron will only become apparent when
$B_{\n[2]}$ approaches $B_{\D}$ by which time this reaction will
be of little importance.

The $\n[2](\p,\gamma)\T$ reaction has been included as a failsafe
mechanism to remove dineutrons since it is exothermic for all
$B_{\n[2]}$. It is not expected to be an important reaction unless
we strongly deviate from the nucleon flow in standard BBN because
the similar, standard BBN reaction $\D(\n,\gamma)\T$ is also
unimportant. We have also estimated its cross section from the
$\n(\p,\gamma)\D$ reaction in \cite{Rupak:1999rk} after modifying
the phase space and spin multiplicity factors using the
expressions discussed above.


\section{The effects of a bound dineutron upon BBN.} \label{sec:BBN+n2 results}

The chief manner in which the dineutron affects BBN is via the
Q-values for the reactions and, from the values quoted earlier, we
can identify three regions of $B_{\n[2]}$ up to the $4\;{\rm MeV}$
limit we considered. They are:
\begin{itemize}
\item $B_{\n[2]} \leq B_{\D}$,
\item $B_{\D} \leq B_{\n[2]} \leq 3.0\;{\rm MeV}$, and
\item $3.0\;{\rm MeV} \leq B_{\n[2]}$.
\end{itemize}
In figure (\ref{fig:dyovery}) we plot the fractional change in the
primordial abundance relative to standard BBN at $\eta=6.14 \times
10^{-10}$ and the three regions we have identified are clearly
visible. Up to $B_{\D}$ there is no discernable change with
$B_{\n[2]}$, over the interval $B_{\D} \leq B_{\n[2]} \leq
3.0\;{\rm MeV}$ the deuterium and helium-3 abundances drop and the
helium-4 abundance rises but after $3.0\;{\rm MeV}$ the helium-3
and helium-4 abundances plateau and the evolution of $Y_{\D}$
changes noticeably. In what follows we shall explain why and how
the dineutron influences BBN in each region.
\begin{figure}[t]
\begin{center}
\epsfxsize=3.4in \epsfbox{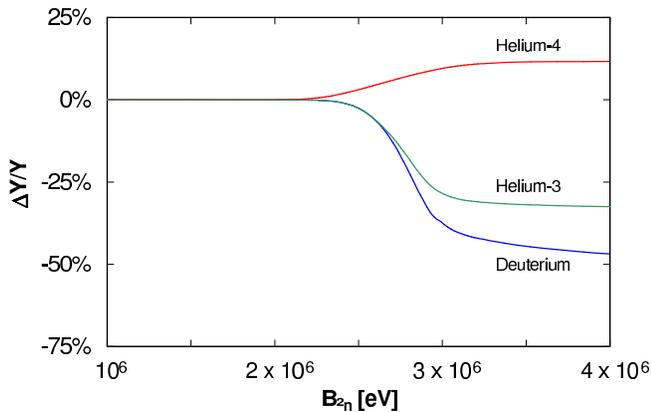} \caption{The fractional
change in the primordial deuterium, helium-3 and helium-4
abundances as a function of the dineutron binding
energy.\label{fig:dyovery} }
\end{center}
\end{figure}


\subsection{$B_{\n[2]} \leq B_{\D}$}

For $B_{\n[2]} \ll B_{\D}$ the dineutron lifetime is of order
$1\;{\rm s}$ so while the decay would appear to be a significant
drain on the dineutron abundance any loss is easily, and rapidly,
replaced from the pool of free neutrons and, consequently, the
dineutron abundance during the second stage of BBN follows its NSE
value:
\begin{equation}
Y_{\n[2]} =
\frac{1}{3}\,\frac{Y_{\n}\,Y_{\D}}{Y_{\p}}\,\exp\,\left(
\frac{B_{\n[2]}-B_{\D}}{T} \right).
\end{equation}
\begin{figure}[b]
\begin{center}
\epsfxsize=3.4in \epsfbox{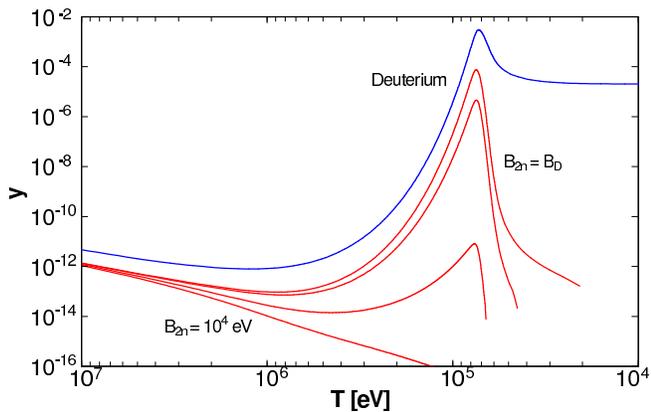} \caption{The evolution of
the dineutron abundance as a function of the temperature for
$B_{\n[2]} \in \{ 10^{4}\;{\rm eV},10^{6}\;{\rm eV},2\times
10^{6}\;{\rm eV},B_{\D} \}$ and the deuteron abundance from
standard BBN. \label{fig:yn2,D vs T.Bn2 < BD} }
\end{center}
\end{figure}
Therefore the effect of increasing the dineutron binding energy is
the same as one expects from NSE, i.e. the abundance builds up at
higher temperature as the binding energy increases. The evolution
of the dineutron abundance as a function of the temperature for
the four cases $B_{\n[2]} \in \{ 10^{4} \, {\rm eV},10^{6} \, {\rm
eV} ,2\times 10^{6} \, {\rm eV},B_{\D} \}$ is shown in figure
(\ref{fig:yn2,D vs T.Bn2 < BD}). In every example in the figure
the dineutron abundance is smaller than the deuteron abundance:
even when $B_{\n[2]} = B_{\D}$ the NSE abundance is smaller than
$Y_{\D}$ due to both the smaller abundance of free neutrons
($Y_{\n}/Y_{\p} \sim 1/6 - 1/7)$ and the spin factor, $1/3$, of
deuterons, combining for a total of $\sim 1/20$. The figure also
makes it quite clear that as $B_{\n[2]}$ approaches $B_{\D}$ the
dineutron starts to possess a considerable abundance at the
transition to BBN proper, approximately the location of the
deuteron peak. And the figure also shows that the position of the
peak dineutron abundance is unaffected by
its binding energy and is coincident with the deuteron peak.
Whatever the dineutron binding energy in this range BBN proper is
still initiated by the deuteron's departure from NSE and the small
amount of dineutrons present at that time is rapidly removed. When
$B_{\n[2]} \ll B_{\D}$ their small abundance means that they never
form a substantial population which could possibly influence the
predictions of BBN but as $B_{\n[2]}$ approaches $B_{\D}$ their
presence at $T_{BBN}$ becomes important.


\subsection{$B_{\D} \leq B_{\n[2]} \leq 3.0\;{\rm MeV}$}

As shown above, the NSE dineutron abundance at $T \sim 100\;{\rm
keV}$ when $B_{\n[2]} = B_{\D}$ is smaller than $Y_{\D}$ by
roughly a factor of $1/20$ so it is not until the dineutron
binding energy has grown to the point where it is capable of
reversing the suppression of its NSE abundance by the
neutron-to-proton ratio and the spin of the deuteron that its
abundance approaches $Y_{D}$ at $T\sim 100\;{\rm keV}$. Roughly
this occurs when
\begin{equation}
\exp\,\left( \frac{B_{\n[2]}-B_{\D}}{T} \right) \sim 20
\end{equation}
i.e. when $B_{\n[2]} \sim 2.5\;{\rm MeV}$. The evolution of the
dineutron and deuteron abundances during this interval for
$B_{\n[2]}$ is shown in figure (\ref{fig:yn2,D vs T.BD < Bn2 < 3})
and confirms that the dineutron abundance surpasses the deuteron
abundance at $B_{\n[2]} \sim 2.5\;{\rm MeV}$.
\begin{figure}[htbp]
\begin{center}
\epsfxsize=3.4in \epsfbox{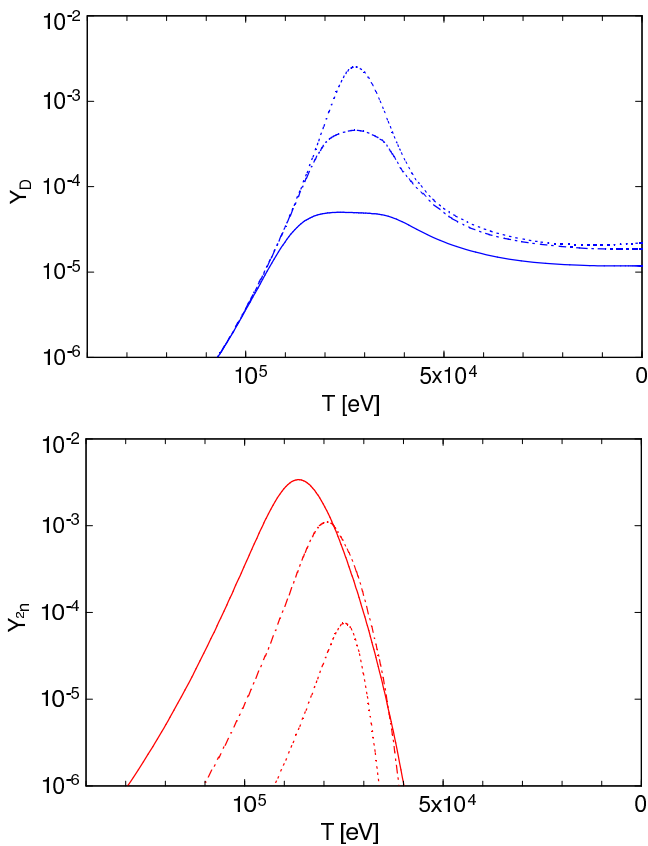} \caption{The
evolution of the deuterium (top panel) and dineutron (bottom
panel) abundances as a function of the temperature for $B_{\n[2]}
\in \{ B_{\D}$ (dotted line), $2.6\;{\rm MeV}$ (dot-dashed line),
$3 \;{\rm MeV}$(solid line)$\}$. \label{fig:yn2,D vs T.BD < Bn2 <
3} }
\end{center}
\end{figure}
The figure also demonstrates a number of new features: the
position of the dineutron peak abundance now moves to higher
temperatures as $B_{\n[2]}$ increases, the peak deuterium
abundance drops and the temperature at which it departs NSE also
shifts to higher temperatures. These effects ripple through to the
triton and helion as shown in figure (\ref{fig:yT,He3 vs T.BD <
Bn2 < 3}).
\begin{figure}[htbp]
\begin{center}
\epsfxsize=3.4in \epsfbox{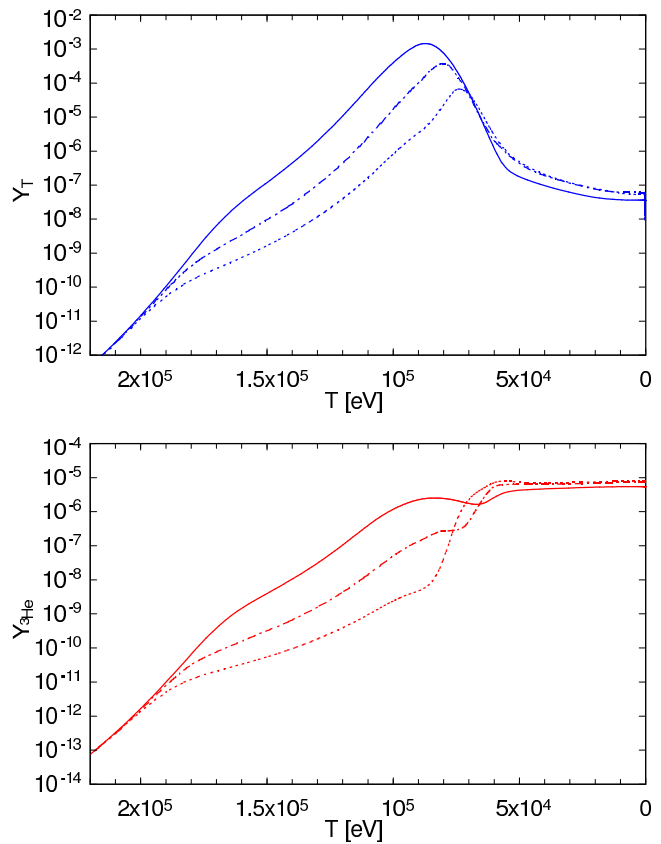} \caption{The
evolution of the tritium (top panel) and helium-3 (bottom panel)
abundances as a function of the temperature for $B_{\n[2]} \in \{
B_{\D}$ (dotted line), $2.6\;{\rm MeV}$ (dot-dashed line), $3
\;{\rm MeV}$ (solid line) $\}$. \label{fig:yT,He3 vs T.BD < Bn2 <
3} }
\end{center}
\end{figure}
As $B_{\n[2]}$ increases tritium and helium-3 depart NSE at lower
temperatures: the change isn't dramatic, by $B_{\n[2]}=3\;{\rm
MeV}$ the two $A=3$ nuclei depart at $T \sim 170\;{\rm keV}$
rather than at around $200\;{\rm keV}$ in standard BBN but the
sensitivity of the NSE abundance to the temperature means that the
abundance of helium-3 and tritium after this point is roughly
$2-3$ orders of magnitude larger. The figure also shows that the
movement of the tritium peak parallels that for dineutrons by
shifting to higher temperatures as $B_{\n[2]}$ increases.

The movements in the abundances of the intermediary nuclei are all
attributable to changes in the flow of the nuclei through the
reaction network. For $B_{\n[2]} \geq B_{\D}$, the reaction
$\D(\n,\p)\n[2]$ is now exothermic and a detailed examination of
the nuclear flow at $B_{\n[2]}=2.6\;{\rm MeV}$ confirms this to be
the source of dineutrons. The flow also indicates the dineutrons
are then chiefly converted to tritons via the reaction
$\n[2](\D,\n)\T$. Both reactions lead to a disruption of the usual
mechanisms that lead to BBN proper. In standard BBN the transition
to BBN proper was due to the removal of deuterons via the two
$\D+\D$ processes but now, because beyond $B_{\n[2]} \sim
2.5\;{\rm MeV}$ the NSE abundance of dineutrons is \emph{larger}
than that of deuterons and because free neutrons are so plentiful
during the second stage of BBN, the departure occurs earlier.
Remarkably both $\D(\D,\n)\He[3]$ and $\D(\D,\p)\T$ are now of
little consequence. The shift in the deuteron's NSE departure
temperature is not large, figure (\ref{fig:yn2,D vs T.BD < Bn2 <
3}) demonstrates this, but, as with \T~ and \He[3], the deuteron's
NSE abundance is very sensitive to the temperature. The
$\n[2](\D,\n)\T$ reaction can also provide sufficient tritium to
keep its abundance in equilibrium until a slightly lower
temperature. The switch in the mechanism that initiates BBN proper
explains many of the features seen in figures (\ref{fig:yn2,D vs
T.BD < Bn2 < 3}) and (\ref{fig:yT,He3 vs T.BD < Bn2 < 3}).

From the nuclear flow we also find that helium-3 no longer plays
an important role in the formation of helium-4. Some helium-3 is
still formed, via $\D(\D,\n)\He[3]$, but this source is suppressed
due to the lack of deuterons; what little helium-3 is created is
processed to tritium by the usual $\He[3](\n,\p)\T$ though
$\He[2](\n[2],\n)\He$ does play a role. The reduced significance
of helium-3 is not reflected in figure (\ref{fig:yT,He3 vs T.BD <
Bn2 < 3}): one must remember that the net rate of formation for
the intermediary \D, \T~ and \He[3] is the small difference
between production and destruction and does not necessarily
indicate the true amount of nucleons flowing through them. Even in
standard BBN the evolution of helium-3 does not resemble that of
tritium because $\He[3](\n,\p)\T$ is so rapid and one can only see
the significance of the helium-3 nucleus by examining the
individual reaction rates \cite{Smith:1992yy}.

Finally, the formation of helium-4 still occurs via $\T(\D,\n)\He$
and, due to the earlier initiation of BBN proper, its final
(primordial) abundance is enhanced. The reaction network is
modified and the bulk of the nucleons now flow through a network
resembling that shown in figure (\ref{flow:2.6MeV}).
\begin{center}
\begin{figure}[htbp]
\setlength{\unitlength}{1.5cm}
\begin{picture}(1.25,2.5)(0,0)
\put(0,2){\framebox(0.5,0.5){\n}}
\put(0.5,2){\makebox(0.25,0.5){+}}
\put(0.75,2){\framebox(0.5,0.5){\D}}
\put(0,1){\framebox(0.5,0.5){\D}}
\put(0.5,1){\makebox(0.25,0.5){+}}
\put(0.75,1){\framebox(0.5,0.5){\n[2]}}
\put(0.75,0){\framebox(0.5,0.5){\T}} \put(1,2){\vector(0,-1){0.5}}
\put(1,1){\vector(0,-1){0.5}}
\end{picture}
\caption{The schematic flow of nuclei through the reaction network
from deuterium to tritium at $B_{\n[2]}=2.6\;{\rm MeV}$. The flow
does not include helium-3. A little of this nucleus is still
produced via $\D(\D,\n)\He[3]$ but the suppressed deuteron
abundance means that the amount is substantially reduced compared
to standard BBN. \label{flow:2.6MeV}}
\end{figure}
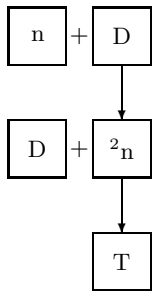
\end{center}

By $B_{\n[2]}=3\;{\rm MeV}$ figure (\ref{fig:yn2,D vs T.BD < Bn2 <
3}) shows that the peak abundance of deuterons has dropped by two
orders of magnitude from that in standard BBN and BBN proper
begins at an even higher temperature. The nuclear flow now
indicates that tritium formation is increasingly dominated by
$\n[2](\p,\gamma)\T$ rather than $\n[2](\D,\n)\T$ and,
interestingly, the dominant source of helium-3 is the mildly
endothermic $\T(\p,\n)\He[3]$ at higher temperatures with a
changeover to $\T(\D,\n[2])\He[3]$ as the Universe cools. This
flow, $\T(\p,\n)\He[3]$, is opposite to that in standard BBN,
$\He[3](\n,\p)\T$.

This second region of $B_{\n[2]}$ is where the presence of
dineutrons really becomes manifest. At $B_{\n[2]} = B_{\D}$ the
dineutron is just starting to influence BBN, by $B_{\n[2]} =
3\;{\rm MeV}$ it has considerably altered the reaction network
that take the free neutrons to helium-4 and led to a significant
change in the mechanism that leads to BBN proper.


\subsection{$3.0\;{\rm MeV} \leq B_{\n[2]}$}

The excitement seen when $B_{\D} \leq B_{\n[2]} \leq 3\;{\rm MeV}$
has largely played out as we enter the third domain of $3\;{\rm
MeV} \leq B_{\n[2]}$ and the shifts in the evolution of the
abundances slows. In this range, between 3 MeV and 4 MeV, there
are no, new, major changes in the nuclear flow because the
Q-values of the most important reactions are now all $\sim \;{\rm
MeV}$. Nevertheless, a detailed study of the reaction network does
show small differences and we discuss those here.

Although further increases in the dineutron binding energy are not
reflected in the position and amplitude of the peak dineutron
abundance,
the temperature at which the dineutron departs NSE shifts to
higher values with $B_{\n[2]}$. In contrast, the reduction
in the amplitude of the peak deuteron abundance seen in figure
(\ref{fig:yn2,D vs T.BD < Bn2 < 3}) becomes less dramatic
and its NSE departure temperature has all but ceased to move as $B_{\n[2]}$
increases.

The behavior of the evolution of these two $A=2$ nuclei reflect
the fact that dineutrons are primarily created from deuterons and
that the dominant dineutron destruction mechanism has switched
from $\n[2](\D,\n)\T$ to $\n[2](\p,\gamma)\T$. Dineutron NSE
departure occurs because of insufficient production from the small
\D~ abundance.
The \n[2] abundance
at lower temperatures is thus controlled by the deuteron. The
reaction $\n[2](\p,\gamma)\T$ is affected by $B_{\n[2]}$ only
through the exit channel phase space which varies as
$\sqrt{B_{\T}-B_{\n[2]}}$. With an abundance controlled by \D~ and
a destruction mechanism that varies only weakly with $B_{\n[2]}$
the height and temperature of the peak dineutron abundance is,
essentially, static. Deuterium departure from NSE occurs when its
abundance opens the $\D(\n,\p)\n[2]$ drain and, again, $B_{\n[2]}$
only enters weakly through the exit channel phase space.

A new feature emerges in this third region of $B_{\n[2]}$. At
$B_{\n[2]}\geq 3.0\;{\rm MeV}$, the Q-value for the reaction
$\He[3](\n[2],\D)\T$ is now negative and the importance of the
helion as an intermediary in the formation of helium-4 rebounds,
though not to the level in standard BBN. The source of helions is
principally the mildly endothermic $\T(\p,\n)\He[3]$ but switches
to $\T(\D,\n[2])\He[3]$ as the Universe cools.

Finally, helium-4 is now chiefly formed by both $\T(\D,\n)\He$ and
$\T(\T,2\n)\He$ in almost equal proportions with minor
contributions from $\T(\p,\gamma)\He$ and $\He[3](\n[2],\n)\He$.
The network has changed slightly and the new flow at
$B_{\n[2]}=3.5\;{\rm MeV}$ is illustrated in figure
(\ref{flow:3.5MeV}).
\begin{center}
\setlength{\unitlength}{1.5cm}
\begin{figure}[htbp]
\begin{picture}(2,1.5)(0,0)
\put(0,1){\framebox(0.5,0.5){\D}}
\put(0.5,1){\makebox(0.25,0.5){,}}
\put(0.75,1){\framebox(0.5,0.5){\T}}
\put(1.25,1){\makebox(0.25,0.5){+}}
\put(1.5,1){\framebox(0.5,0.5){\T}}
\put(1.5,0){\framebox(0.5,0.5){\He}}
\put(1.75,1){\vector(0,-1){0.5}}
\end{picture}
\caption{The diagrammatic flow of nuclei through the reaction
network from tritium onwards at $B_{\n[2]}=3.5\;{\rm MeV}$.
\label{flow:3.5MeV}}
\end{figure}
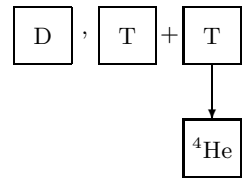
\end{center}


\section{Are these predictions robust?} \label{sec:MC}

While we have strived to estimate the various dineutron cross
sections by basing them upon simple physical assumptions, it is a
worthwhile task to see how robust the effects we have described
are. If we know (or assume) the distribution for the error, either
in the cross-section or the thermally averaged rate, then one can
compute the error matrix as in Fiorentini {\it et al.}
\cite{FLSV1998} and Cuoco {\it et al.} \cite{Cetal2003} though
this approach cannot recover higher moments of the abundance
distribution. An alternative is to construct the distribution of
the abundances by sampling the distribution of the errors in a
Monte Carlo analysis. This technique was used by Krauss \&
Romanelli \cite{KR1990}, Smith, Kawano \& Malaney
\cite{Smith:1992yy}, Krauss \& Kernan \cite{KK1995} and, more
recently, Nollett and Burles \cite{NB2000} and is the technique we
shall use. To this end we introduced random multiplicative factors
for all our dineutron reaction cross sections with the exception
of $\n[2] \leftrightarrow \D$, the most reliably estimated. These
factors were chosen from a probability distribution limited to the
range between 1/5 and 5 and weighted such that there was equal
probability either side of 1. The baryon-to-photon ratio was fixed
at $\eta=6.14\times 10^{-10}$.
\begin{figure}[htbp]
\begin{center}
\epsfxsize=3.4in \epsfbox{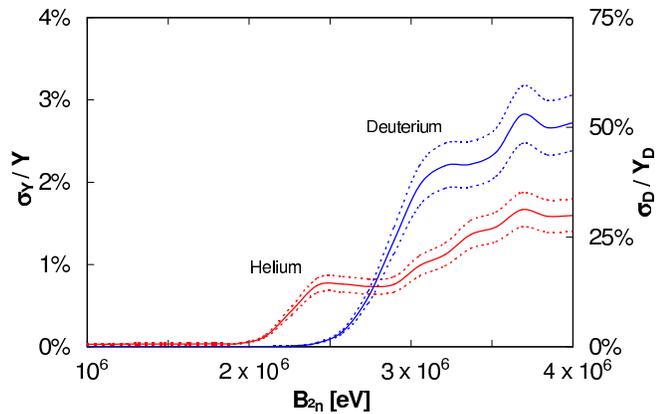} \caption{The root mean square
deviation divided by the mean plus the $\pm 3\sigma$ error for
deuterium and helium-4 as a function of the dineutron binding
energy. Below $B_{\n[2]} \sim 2\;{\rm MeV}$ neither nucleus
exhibits any spread due to the uncertainty in the dineutron
reaction rates. Above this value the spread in the deuterium
results becomes large while the results for helium remain smaller
than $\sim 2\%$ for all value of $B_{\n[2]}$. \label{fig:YDrms} }
\end{center}
\end{figure}
In figure (\ref{fig:YDrms}) we plot the root mean square
deviations for deuterium and helium-4 and the three regimes for
the dineutron binding energy are clearly visible in both curves.
The increasing spread for both reflects the increasing dominance
of dineutron reactions in the formation of helium-4. The figure
shows that up to $B_{\n[2]} \sim 2\;{\rm MeV}$ the errors in the
reactions involving dineutrons do not introduce any error into the
predicted abundances, illustrating again the insignificance of
dineutrons in BBN when their binding energy is smaller than this
value. Above $B_{\n[2]} \sim 2\;{\rm MeV}$ the curves begin to
deviate from zero but note two important features: first, the
spread in helium-4 is small for all values of $B_{\n[2]}$, and
second, the spread in the predicted abundance of deuterium does
not exceed $\pm100\%$. The small spread in the helium-4 curves
indicate that we can reliably predict the abundance of helium-4
even if dineutrons have significantly impacted BBN, while the fact
that the deuterium curves do not exceed $100\%$ shows that at
least the \emph{direction} of the change is known even if the
exact abundance is not. We have not shown the rms spread for
helium-3 because it exceeded $\pm100\%$.

Note also that the range in the abundances is significantly
smaller than the range we permitted for the random factors. This
may seem remarkable since the primordial abundance of an
intermediary nuclei, such as deuterium, involves a fierce
competition between its production and destruction and hence the
adopted cross sections. While dineutrons may have led to
significant departures in the flow of nuclei compared to standard
BBN, large variations in the dineutron reaction rates do not
translate into equally large spreads in the abundances of the
intermediaries.

\begin{figure}[htbp]
\begin{center}
\epsfxsize=3.4in \epsfbox{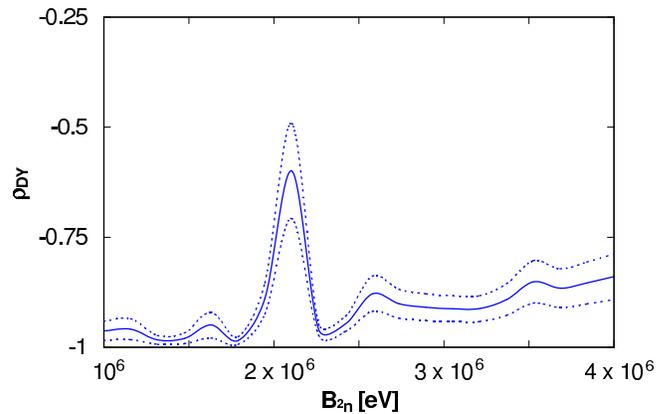} \caption{The
correlation between the deuterium abundance and helium-4 mass
fraction, $\rho_{DY}$, and its $\pm 3\sigma$ error, as a function
of the dineutron binding energy. Below $B_{\n[2]} \sim 2\;{\rm
MeV}$ the correlation is close to $-100\%$ while after $B_{\n[2]}
\sim 3\;{\rm MeV}$ the correlation has softened to $\sim -80\%$.
The large peak at $B_{\n[2]} \sim 2\;{\rm MeV}$ corresponds to the
point where the change in the flow of nuclei through the reaction
network occurs. \label{fig:YDcorrelation} }
\end{center}
\end{figure}
We also found that there was a significant anti-correlations in
the results as shown in figure (\ref{fig:YDcorrelation}). For
$B_{\n[2]} \lesssim 2\;{\rm MeV}$ the correlation coefficient was
almost exactly $-1$ while over the interval $2.5\;{\rm MeV}
\lesssim B_{\n[2]} \lesssim 3\;{\rm MeV}$ the anti-correlation
softened slightly to $\sim -0.8$. As a consequence of this
correlation the covariance matrix, $V_{R}$, describing the error
in the predictions arising from the uncertainties in the
dineutron's reaction rates, contains off-diagonal pieces.

\begin{figure}[htbp]
\begin{center}
\epsfxsize=3.4in \epsfbox{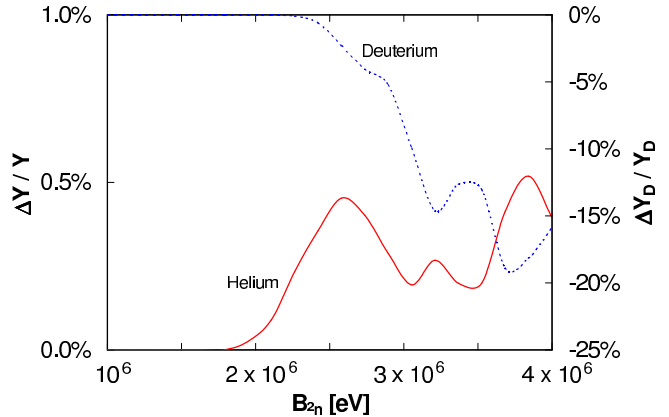} \caption{The fractional
difference between the mean from the Monte Carlo simulation and
the result with no random factors. The solid line is for helium-4,
the dotted is for deuterium. \label{fig:drift} }
\end{center}
\end{figure}
We have delayed until last the difference we find in the mean from
the sample and the primordial abundances we derive with no random
factors as shown in figure (\ref{fig:dyovery}). The fact that a
discrepancy arises is not surprising given the non-linearity of
BBN. The errors (which are functions of $B_{\n[2]}$) for helium-4
is $\lesssim 0.5\%$ but for deuterium it reaches $\lesssim 20\%$
and helium-3 fared even worse with a discrepancy between the mean
and no random factors approaching $30\%$. The fractional
difference between the mean and the result with no random factors
are shown in figure (\ref{fig:drift}) for deuterium and helium-4.
Although seemingly large, this error is smaller than the
statistical fluctuations for all three nuclei (seen in figure
(\ref{fig:YDrms}) for deuterium and helium-4) but not
significantly so and we must include it in the total error for the
predictions as a systematic. The covariance matrix for the
systematic error is denoted by $V_{S}$.

With means, ${\bf \bar{Y}}$, and variance, $V_{T}=V_{R}+V_{S}$,
that are functions of $B_{\n[2]}$ we approximate the distributions
in the predictions as Gaussians
\begin{eqnarray}
& & P({\bf Y}|B_{\n[2]}) = \frac{1}{\sqrt{2\pi\,|V_{T}|}} \nonumber \\
& & \times \exp\left[ -\frac{1}{2}\,({\bf Y}-{\bf
\bar{Y}})^{T}\,V_{T}^{-1}\,({\bf Y}-{\bf \bar{Y}}) \right].
\label{eq:gaussian}
\end{eqnarray}
We have checked the validity of this approximation to the spread
in the results by performing a Kolmogorov test at each value of
$B_{\n[2]}$ we set. The spread in the predictions for helium-3,
which we have not shown, did not pass this due to a significant
kurtosis.


\section{A degeneracy with $\eta$?} \label{sec:eta}

So far we have restricted our discussions to a fixed value of the
baryon-to-photon ratio, $\eta$, but if we allow this parameter to
also vary we may well end up with a degeneracy that makes it
difficult to establish the presence of a bound dineutron.
\begin{figure}[htbp]
\begin{center}
\epsfxsize=3.4in \epsfbox{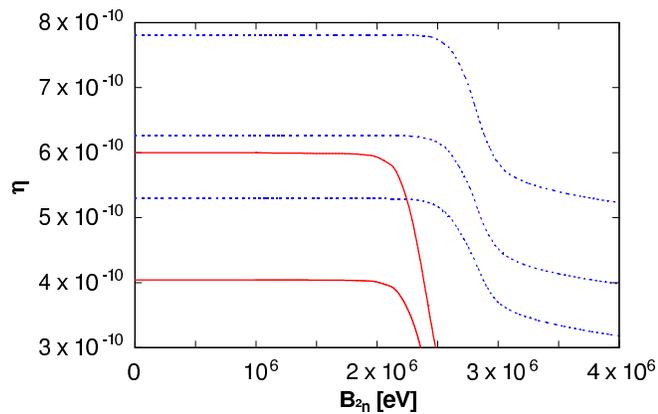} \caption{Helium-4
iso-mass fraction (solid) and Deuterium iso-abundance (dashed)
contours in the $B_{\n[2]}-\eta$ plane. From top to bottom, the
helium-4 contours are $Y=0.248$ and $Y=0.244$ and the deuterium
abundances are $\D/H= 1.8\times 10^{-5}$, $\D/H= 2.6\times
10^{-5}$ and $\D/H= 3.4\times 10^{-5}$. \label{fig:yD,YvsBn2,eta}
}
\end{center}
\end{figure}
To determine if this occurs, we show in figure
(\ref{fig:yD,YvsBn2,eta}) iso-abundance and iso-mass fraction
contours for deuterium and helium respectively as a function of
$B_{\n[2]}$ and $\eta$.  If we examine the most robust prediction
of our modified BBN, the increase in $Y$, then we can quite easily
mask this effect by \emph{lowering} $\eta$ as shown in the figure.
The decrease in $\eta$ required to offset the increase in the
primordial mass fraction as $B_{\n[2]}$ increases is considerable
so that above $B_{\n[2]} \sim 2.5\;{\rm MeV}$ all values of $\eta$
in the $3 \times 10^{-10}$ to $8 \times 10^{-10}$ range plotted
yield a helium-4 mass fraction above $0.248$. For deuterium, lower
values of $\eta$ lead to increases in the final abundance so that
they too can mitigate the decrease in $Y_{\D}$ as $B_{\n[2]}$
increases. But the figure shows that the decrease in $\eta$
required for deuterium is nowhere near as large as that required
for helium-4, so, while there is an anticorrelation of $B_{\n[2]}$
and $\eta$ for both helium-4 and deuterium, there is no degeneracy
for both simultaneously. If we considered each species separately
then the degeneracy could, instead, be broken by using the CMB
since lowering $\eta_{BBN}$ will lead to a discrepancy with
$\eta_{CMB}$.


\section{An upper limit for $B_{\n[2]}$} \label{sec:limits}

It's finally time to derive an upper limit to $B_{\n[2]}$ based on
the compatibility with observations. The primordial abundance,
$\D/{\rm H}$, of deuterium is taken to be $\D/{\rm H} = ( 2.6 \pm
0.4 )\times 10^{-5}$ \cite{Betal2003a} while we will consider both
the Olive, Steigman and Walker \cite{OSW2000} value of $Y_{OSW} =
0.238 \pm 0.005$ and the Izotov and Thuan value of $Y_{IT}=0.244
\pm 0.002$ \cite{ITL1997,IT1998} for the helium-4 mass fraction
$Y$. The exact primordial abundances remain a topic of debate with
two, largely incompatible, determinations for the helium mass
fraction \cite{ITL1997,IT1998,OS1995,OSS1997,FO1998} and excessive
scatter in the measurements of deuterium
\cite{Ketal2003,Betal2003a} but these two nuclei still represent
the best probes of BBN.

The error in the observed helium-4 mass fractions are 2\% for
$Y_{OSW}$, 1\% for $Y_{IT}$ which compares well with the spread in
the predicted mass fraction plotted in figure (\ref{fig:YDrms}).
We can integrate over the possible values for the prediction, at a
fixed $\eta$ and $B_{\n[2]}$, using the distributions found
earlier, and find
\begin{eqnarray}
& & {\cal L}(\eta,B_{\n[2]}|{\bf \hat{Y}}) = \frac{1}{\sqrt{2\pi\,|V|}} \nonumber \\
& & \times \exp\left[ -\frac{1}{2}\,({\bf \hat{Y}}-{\bf
Y})^{T}\,V^{-1}\,({\bf \hat{Y}}-{\bf Y}) \right],
\end{eqnarray}
where ${\bf \hat{Y}}$ denotes the vector whose elements are the
observations and ${\bf Y}$ the vector whose elements are the
predictions and $V$ is the covariance matrix - the sum of the
(diagonal) covariance matrix for the observations, $V_{O}$, and
the two covariance matrices $V_{R}$ and $V_{S}$. Contours of the
likelihood are shown in figure (\ref{fig:contours}) which shows
that the use of deuterium and helium-4 breaks the degeneracy seen
in each separately, and, indeed, the limits to the dineutron
binding energy are independent of the baryon-to-photon ratio.
\begin{figure}[htbp]
\begin{center}
\epsfxsize=3.4in \epsfbox{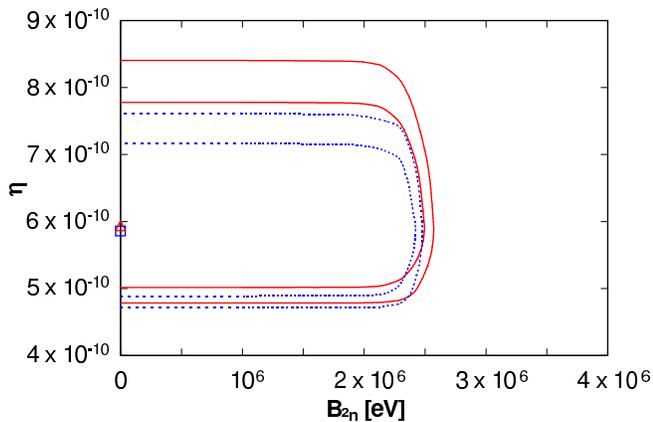} \caption{The 95\% and 99\%
confidence contours using the Olive, Steigman and Walker (solid
curves) and Izotov and Thuan (dashed curves) helium-4 mass
fractions observations and the Barger \emph{et al.}
\cite{Betal2003a} primordial deuterium abundance. The best fit
points, both at $B_{\n[2]}=0$, are denoted by the triangle for
OSW, the square for IT. \label{fig:contours} }
\end{center}
\end{figure}

Marginalizing over the baryon-to-photon ratio, $\eta$, we obtain
the results shown in figure (\ref{fig:marginalized}). The two
upper limits for $B_{\n[2]}$ are not greatly influenced by the
different primordial helium-4 mass fractions found in the
literature.
\begin{figure}[htbp]
\begin{center}
\epsfxsize=3.4in \epsfbox{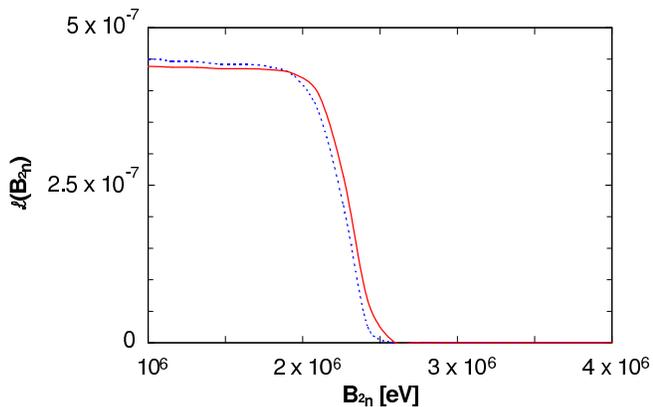} \caption{The
marginalized likelihood function for $B_{\n[2]}$ using the Olive,
Steigman and Walker (solid curve) and Izotov and Thuan (dashed
curve) helium-4 mass fractions observations and the Barger
\emph{et al.} \cite{Betal2003a} primordial deuterium abundance.
\label{fig:marginalized} }
\end{center}
\end{figure}


\section{Summary and Conclusions}

We have examined the role that a stable dineutron may play in BBN
as a function of its binding energy, $B_{\n[2]}$, up to $4\;{\rm
MeV}$. We have estimated the important new reactions that enter
into the BBN reaction network and examined, in detail, the change
in the nucleon flow. We find that the range $0 \leq B_{\n[2]} \leq
4\;{\rm MeV}$ can be subdivided into three regions: $B_{\n[2]}
\leq B_{\D}$, $B_{D} \leq B_{\n[2]} \leq 3\;{\rm MeV}$ and
$3\;{\rm MeV} \leq B_{\n[2]}$. The boundaries are due to the sign
inversion of the $\n[2](\p,\n)\D$ and $\He[3](\n[2],\D)\T$
Q-values respectively. Below $B_{\D}$ the dineutron has little
effect upon BBN but as we increased $B_{\n[2]}$ beyond this value
it began to interfere with the nucleon flow between deuterons and
tritium. The helium-4 mass fraction increased by $\sim 10\%$ and
the deuterium abundance dropped by $\sim 40\%$. Above $3\;{\rm
MeV}$ the nucleon flow settled into a new pattern, the helium-4
mass fraction plateaued to new level and the rate at which the
deuterium abundance decreases with $B_{\n[2]}$ slowed. A small
change to the nucleon flow from tritium to helium-4 was seen.

We estimated the error in the predictions by sampling the
distribution of abundances when the dineutron reactions were
multiplied by random factors and found that the prediction of an
increase in the helium-4 mass fraction and decrease in deuteron
abundance were reliable. The degeneracy between $B_{\n[2]}$ and
the baryon-to-photon ratio, $\eta$, that occurs for $Y$ and
$\D/\Hy$ separately was broken when both were considered
simultaneously. We then constructed the 2-D likelihood function
${\cal L}(\eta,B_{\n[2]}|Y,D/H)$ by using both the OSW
\cite{OSW2000} and IT \cite{ITL1997,IT1998} helium-4 mass
fractions and the Barger {\it et al.} \cite{Betal2003a} deuterium
abundance and then marginalized over $\eta$ to derive the
likelihood distribution for $B_{\n[2]}$. We found that dineutron
binding energies above $\sim 2.5\;{\rm MeV}$ could not be
accommodated by BBN within both allowed ranges of Y

Throughout our calculation we have only permitted the variation of
the dineutron binding energy. In reality, whatever the source of
the stability of the dineutron, naively the binding energies of
the other nuclei should also change. The range in $B_{\n[2]}$ we
investigated is much larger than the range in $B_{\D}$ that
Kneller \& McLaughlin permitted but we notice that the increase in
$Y$ and decrease in $\D/\Hy[]$ seen there when $B_{\D}$ increased is
also mimicked by the dineutron. Thus if $B_{\D}$ and $B_{\n[2]}$
increase in tandem then the stability of the dineutron cannot
reverse the increase in $Y$ whatever the relative magnitudes of
the two binding energies and even though dineutrons may alter the
nucleon flow. If these two binding energies change in opposite
senses then the situation is more confused: the effect of a
decrease in $B_{\D}$ is the immediate decrease in the helium-4
mass fraction which may not be reversed by the presence of a
stable dineutron unless $B_{\n[2]} \gtrsim B_{\D}$. A better
understanding of the net effect, and a more concrete limit to the
permitted variation of both the deuteron and dineutron binding
energies, could be derived if the relationship between them were
known and we were not forced to have to treat each as independent.
Furthermore, if such a calculation shows that the predicted
helium-4 mass fraction can be consistent with observations,
without changing the compatibility of $\eta_{BBN}$ and
$\eta_{CMB}$ and the deuterium abundance greatly, then an
examination of the effects upon the primordial yield of lithium-7,
which, in standard BBN, is overproduced theoretically, would be in
order. Although we have not included any dineutron reaction
involving nuclei with mass above $A=4$, one can speculate that the
omitted reaction $\Be(\n[2],\n\alpha)\He$ could play a significant
role since beryllium-7 (before it decays to \Li) is the chief
component of the primordial $A=7$ isobar yield at $\eta \sim
6\times 10^{-10}$.

In this paper we have shown that the dineutron can become bound at
a level up to that of the deuteron without disrupting the standard
nuclear flow in BBN or significantly altering predicting BBN
abundance yields. Beyond that, changes to the nuclear flow and to
predicted abundance yields appear. It is important to note that we
have allowed only one parameter, the dineutron binding energy, to
vary. Further work on the interdependence of cross sections and
binding energies in nuclear theory would be required to reduce the
errors presented here and to make a more concrete connection with
the underlying fundamental constants.


\vspace{1.cm} \acknowledgements The authors would like to thank
Eric Braaten for useful discussions. This work was supported by
the U.S. Department of Energy under grant DE-FG02-02ER41216.




\newpage

\begin{thebibliography}{99}

\bibitem{S1998} D.~N.~Schramm, Space Science Reviews, {\bf 84}, 3 (1998)
\bibitem{VFCC2003} E.~Vangioni-Flam, A.~Coc and M.~Cass\'{e}, Nuc. Phys. A, {\bf 718}, 389 (2003)
\bibitem{SSG1977} G.~Steigman, D.~N.~Schramm and J.~R.~Gunn, Phys. Let. B, {\bf 66}, 202 (1997)
\bibitem{YSSR1979} J.~Yang, D.~N.~Schramm, G.~Steigman and R.~T.~Rood, Ap. J., {\bf 227}, 697 (1979)
\bibitem{LSV1999} E.~Lisi, S.~Sarkar and F.~L.~Villante, Phys. Rev. D,{\bf 59}, 123520 (1999)
\bibitem{KSSW2001} J.~P.~Kneller \emph{et al.}, Phys. Rev. D, {\bf 64}, 123506 (2001)
\bibitem{BY1977} G.~Beaudet and A.~Yahil, Ap. J., {\bf 218}, 253 (1977)
\bibitem{WFH1967} R.~V.~Wagoner, W.~A.~Fowler and F.~Hoyle, Ap. J., {\bf 148}, 3 (1967)
\bibitem{KS1992} H.-S.~Kang and G.~Steigman, Nucl. Phys. B, {\bf 372}, 494 (1992)
\bibitem{KKS1997} K.~Kohri, M.~Kawasaki and K.~Sato, Ap. J., {\bf 490}, 72 (1997)
\bibitem{OSTW1991} K.~A.~Olive \emph{et al.}, Phys. Lett. B, {\bf 265}, 239 (1991)
\bibitem{S1983} R.~J.~Scherrer, M. N. R. A. S., {\bf 205}, 683 (1983)
\bibitem{YB1976} A.~Yahil and G.~Beaudet, A. and A., {\bf 206}, 415 (1976)
\bibitem{Betal2003a} V.~Barger \emph{et al.}, Phys. Lett. B, {\bf 566}, 8 (2003)
\bibitem{Betal2003b} V.~Barger \emph{et al.}, Phys. Lett. B, {\bf 569}, 123 (2003)
\bibitem{SDT1993} A.~Serna and R.~Dominguez-Tenreiro, Phys. Rev. D, {\bf 48}, 1591 (1993)
\bibitem{SDTY1992} A.~Serna, R.~Dominguez-Tenreiro and G.~Yepes, Ap. J., {\bf 391}, 433 (1992)
\bibitem{BHM2001} R.~Bean, S.~H.~Hansen and A.~Melchiorri, Phys. Rev. D, {\bf 64}, 103508 (2001)

\bibitem{KS2003} J.P.~Kneller and G.~Steigman, Phys. Rev. D, {\bf 67}, 063501 (2003)
\bibitem{CSS2001} X.~Chen, R.~J.~Scherrer and G.~Steigman, Phys. Rev. D, {\bf 63}, 123504 (2001)
\bibitem{F2000} R.~Foot, Phys. Rev. D, {\bf 61}, 023516 (2000)
\bibitem{LS2001} C.~Lunardini and A.~Y.~Smirnov, Phys. Rev. D, {\bf 64}, 073006 (2001)
\bibitem{Detal2002} A.~D.~Dolgov \emph{et al.}, Nucl.\ Phys.\ B {\bf 632}, 363 (2002)
\bibitem{W2002} Y.~Y.~Wong, Phys. Rev. D, {\bf 66}, 025015 (2002)
\bibitem{ABB2002} K.~N.~Abazajian, J.~F.~Beacom and N.~F.~Bell, Phys. Rev. D, {\bf 66}, 013008 (2002)
\bibitem{KSW2000} M.~Kaplinghat, G.~Steigman and T.~P.~Walker, Phys. Rev. D, {\bf 61}, 103507 (2000)
\bibitem{KS1982} E.~W.~Kolb and R.~J.~Scherrer, Phys. Rev. D, {\bf 25}, 1481 (1982)
\bibitem{SSF1993} X.~Shi, D.~N.~Schramm and B.~D.~Fields, Phys. Rev. D, {\bf 48}, 2563 (1993)
\bibitem{SFA1999} X.~Shi, G.~M.~Fuller and K.~Abazajian, Phys. Rev. D, {\bf 60}, 063002 (1999)

\bibitem{Bergstrom:1999wm} L.~Bergstrom, S.~Iguri and H.~Rubinstein, Phys.\ Rev.\ D, {\bf 60}, 045005 (1999)
[arXiv:astro-ph/9902157]

\bibitem{Aetal01} P.P.~Avelino \emph{et al.}, Phys. Rev. D {\bf 64}, 103505 (2001)
\bibitem{Nollett:2002da} K.~M.~Nollett and R.~E.~Lopez, Phys.\ Rev.\ D, {\bf 66}, 063507 (2002)
[arXiv:astro-ph/0204325]

\bibitem{Yoo:2002vw} J.~J.~Yoo and R.~J.~Scherrer,
[arXiv:astro-ph/0211545]
\bibitem{Flambaum:2002de} V.~V.~Flambaum and E.~V.~Shuryak, Phys.\ Rev.\ D, {\bf 65}, 103503 (2002)
[arXiv:hep-ph/0201303]

\bibitem{Flambaum:2002wq} V.~V.~Flambaum and E.~V.~Shuryak,
[arXiv:hep-ph/0212403]
\bibitem{KM2003} J.P.~Kneller and G.C.~McLaughlin, to be published in PRD
\bibitem{Webb:1998cq} J.~K.~Webb {\it et al.}, Phys.\ Rev.\ Lett.\,  {\bf 82}, 884 (1999)
[arXiv:astro-ph/9803165]
\bibitem{Webb:2000mn} J.~K.~Webb {\it et al.}, Phys.\ Rev.\ Lett.\,  {\bf 87}, 091301 (2001)
[arXiv:astro-ph/0012539]
\bibitem{Bahcall:2003rh} J.~N.~Bahcall, C.~L.~Steinhardt and D.~Schlegel,
[arXiv:astro-ph/0301507]

\bibitem{PC54} R.~H.~Phillips and K.~M.~Crowe, Phys. Rev., {\bf 96}, 484 (1954)
\bibitem{IKPS61} K.~Ilakovac {\it et al.}, Phys. Rev., {\bf 124}, 1923 (1961)
\bibitem{GGS1987} W.~R.~Gibbs, B.~F.~Gibson and G.~J.~Stephenson, Phys. Rev. C, {\bf 11}, 90 (1975)
\bibitem{Setal87} O.~Schori {\it et al.}, Phys. Rev. C, {\bf 35}, 2252 (1987)
\bibitem{CH53} B.~L.~Cohen and T.~H.~Handley, , Phys. Rev., {\bf 92}, 101 (1953)
\bibitem{Betal1985a} O.~V.~Bochkarev {\it et al.}, JETP Letters, {\bf 42}, 374 (1985)
\bibitem{Betal1985b} O.~V.~Bochkarev {\it et al.}, JETP Letters, {\bf 42}, 377 (1985)
\bibitem{SP1991} K.~K.~Seth and B.~Parker, Phys. Rev. Letts., {\bf 66}, 2448 (1991)
\bibitem{Metal2003} F.~M.~Marques {\it et al.}, Phys. Rev. C, {\bf 65}, 044006 (2003)

\bibitem{Beane:2002vq}
S.~R.~Beane and M.~J.~Savage,
Nucl.\ Phys.\ A, {\bf 713}, 148 (2003) [arXiv:hep-ph/0206113]

\bibitem{Beane:2002xf}
S.~R.~Beane and M.~J.~Savage,
Nucl.\ Phys.\ A, {\bf 717}, 91 (2003) [arXiv:nucl-th/0208021]

\bibitem{Smith:1992yy}
M.~S.~Smith, L.~H.~Kawano and R.~A.~Malaney,
Astrophys.\ J.\ Supp.\ Series,  {\bf 85}, 219 (1993)

\bibitem{braaten}
E.~Braaten and H.~W.~Hammer,
Phys.\ Rev.\ Lett.\  {\bf 91}, 102002 (2003)
[arXiv:nucl-th/0303038].

\bibitem{FCZ1967} W.~A.~Fowler, G.~R.~Caughlan and B.~A.~Zimmerman, Ann. Rev. Astron. Astro., {\bf 5}, 525 (1967)
\bibitem{Gamow} G.~Gamow, `Structure of atomic nuclei and nuclear transformations, being a second edition of Constitution of atomic nuclei and radioactivity', Oxford, Clarendon Press, (1937).
\bibitem{Bethe} H.~Bethe, Rev. Mod. Phys., vol 9, 69 (1937)
\bibitem{Clayton83} D.D.~Clayton, `Principles of stellar evolution and nucleosynthesis', University of Chicago Press, (1983)
\bibitem{Cetal93} G.S.~Chulick \emph{et al.}, Nucl. Phys. A, {\bf 551}, 255 (1993)
\bibitem{endf} http://www.nndc.bnl.gov/nndc/endf/
\bibitem{TUNL} D.R.~Tilley \emph{et al.}, Nucl. Phys. A, {\bf 708}, 3  (2002)
\\(see also http://www.tunl.duke.edu/nucldata/HTML/ HTML\_Project.shtml)

\bibitem{Rupak:1999rk}
G.~Rupak,
Nucl.\ Phys.\ A, {\bf 678}, 405 (2000) [arXiv:nucl-th/9911018]

\bibitem{Cetal2003} A.~Cuoco {\it et al.},
[arXiv:astro-ph/0307213]

\bibitem{FLSV1998} G.~Fiorentini, E.~Lisi, S.~Sarkar and F.~L.~Villante, Phys. Rev. D, {\bf 58}, 063506 (1998)
\bibitem{KR1990} L.~M.~Krauss and P.~Romanelli, Astrophys. J., {\bf 358}, 47 (1990)
\bibitem{KK1995} L.~M.~Krauss and P.~Kernan, Phys. Letts., {\bf B347}, 347 (1995)
\bibitem{NB2000} K.~M.~Nollett, S.~Burles, Phys.Rev. D, {\bf 61} 123505 (2000)
\bibitem{BBF1989} J.~Bernstein, L.~S.~Brown and G.~Feinberg, Rev. Mod. Phys. {\bf 61}, 25 (1989)
\bibitem{S1996} S.~Sarkar, 1996, Rep. Prog. Phys., {\bf 59}, 1493 (1996)
\bibitem{OSW2000} K.~A.~Olive, G.~Steigman and T.~P.~Walker, Phys. Rep., {\bf 333}, 389 (2000)
\bibitem{ITL1997} Y.~I.~Izotov, T.~X.~Thuan and V.~A.~Lipovetsky, ApJSS., {\bf 108}, 1 (1997)
\bibitem{IT1998} Y.~I.~Izotov and T.~X.~Thuan, ApJ., {\bf 500}, 188 (1998)
\bibitem{OS1995} K.~A.~Olive and G.~Steigman, ApJSS., {\bf 97}, 49 (1995)
\bibitem{OSS1997} K.~A.~Olive, E.~D.~Skillman and G.~Steigman, ApJ., {\bf 483}, 788 (1997)
\bibitem{FO1998} B.~D.~Fields and K.~Olive, ApJ, {\bf 506}, 177 (1998)
\bibitem{Ketal2003} D.~Kirkman \emph{et al.},
[arXiv:astro-ph/0302006]

\end{thebibliography}
\end{document}